\documentclass[
reprint,
superscriptaddress,
 amsmath,amssymb,
 aps,
prc,
floatfix,
showkey,
]{revtex4-2}

\usepackage{graphicx}% Include figure files
\usepackage{dcolumn}% Align table columns on decimal point
\usepackage{bm}% bold math
\usepackage[colorlinks=true,linkcolor=blue,citecolor=blue,urlcolor=blue]{hyperref}
\usepackage[mathlines]{lineno}% Enable numbering of text and display math
\usepackage{url}
\usepackage{booktabs}
\usepackage{tabularx}
\usepackage{array}
\newlength{\singlefigwidth}
\begin{document}

% \preprint{APS/123-QED}
    \title{Dynamical selection of fragment shell effects in spontaneous fission of $^{240}$Pu, $^{232}$Th, and $^{264}$Fm}

\author{Qiafeng Chen}%
% \email{chenqf59@mail2.sysu.edu.cn}%
\affiliation{%
Sino-French Institute of Nuclear Engineering and Technology, Sun Yat-sen University, Zhuhai 519082, China%
}%

\author{Fuchang Gu}%
% \email{gufch3@mail2.sysu.edu.cn}%
\affiliation{%
Sino-French Institute of Nuclear Engineering and Technology, Sun Yat-sen University, Zhuhai 519082, China%
}%

\author{Yingge Huang}%
% \email{huangyg27@mail2.sysu.edu.cn}%
\affiliation{%
Sino-French Institute of Nuclear Engineering and Technology, Sun Yat-sen University, Zhuhai 519082, China%
}%

\author{Erxi Xiao}%
% \email{xiaoerx@mail2.sysu.edu.cn}%
\affiliation{%
Sino-French Institute of Nuclear Engineering and Technology, Sun Yat-sen University, Zhuhai 519082, China%
}%

\author{Yinu Zhang}%
\email{zhangyn226@mail.sysu.edu.cn}%
\affiliation{%
Sino-French Institute of Nuclear Engineering and Technology, Sun Yat-sen University, Zhuhai 519082, China%
}%

\author{Jun Su}%
\email{sujun3@mail.sysu.edu.cn}%
\affiliation{%
Sino-French Institute of Nuclear Engineering and Technology, Sun Yat-sen University, Zhuhai 519082, China%
}%
\affiliation{%
Key Laboratory of Nuclear Data, China Institute of Atomic Energy, Beijing 102413, China%
}%

\begin{abstract}
\begin{description}
\item[Background]
Fragment shell effects are central to the formation of spontaneous-fission (SF) mass yields. 
However, it remains unclear which configurations favored by fragment shell effects are dynamically selected and ultimately appear as peaks in the SF yields.

\item[Purpose]
To clarify, by comparing $^{240}$Pu, $^{232}$Th, and $^{264}$Fm, how the potential energy surface (PES) topology and the fission dynamics in the classically allowed region select the scission configurations in which fragment shell stabilization is reflected in the SF yield peaks.
  
\item[Method]
A two-step microscopic framework is employed. The PES and collective inertia are obtained from constrained Hartree--Fock--Bogoliubov (HFB) calculations. The first step, from the inner to the outer turning points, is treated within the Wentzel--Kramers--Brillouin (WKB) approximation along the least-action path. The second step, from the outer turning points to scission in the classically allowed region, is described by Langevin dynamics. Fragment shell effects are analyzed through fragment PES and smoothed level density (SLD) indicators evaluated for representative even--even fragment pairs selected from the Langevin scission configurations.

\item[Results]

The calculated yields display a two-component asymmetric pattern in $^{240}$Pu, a less distinctly separated asymmetric pattern in $^{232}$Th, and a symmetric peak in $^{264}$Fm.
The fragment shell analysis shows that the larger yield components coincide with a coherent overlap among dynamically populated scission configurations, favorable fragment PES regions, and regions of low neutron and/or proton level density near the Fermi surface.
This overlap is most evident in the dominant asymmetric components of $^{240}$Pu and $^{232}$Th and in the symmetric component of $^{264}$Fm, whereas the subdominant components show weaker or more diffuse overlap.

\item[Conclusion] 

SF yields reflect shell-favored fragment configurations that are made accessible by the PES topology and populated by the Langevin dynamics.
Enhanced yields then arise in shell-stabilized channels, where shell gaps provide additional binding and favor the fragment configurations energetically.
Proton shell effects play a key role in heavy and light fragments, while neutron shell effects provide additional stabilization.

\end{description}
\end{abstract}

\maketitle

\section{Introduction}

Nuclear fission is a fundamental decay mode, and SF provides an extreme manifestation of many-body quantum tunneling~\cite{schmidt2018review,sadhukhan2020microscopic}. SF half-lives, fragment mass and charge yields are important for superheavy element stability~\cite{staszczak2013spontaneous,sanchez2026accurate}, heavy element production in the $r$-process~\cite{martinez2007role}, and plutonium safeguards applications~\cite{croft2021review}. 
Fragment mass and charge distributions are key fission observables as they directly reflect the interplay between collective dynamics and shell effects~\cite{schunck2016microscopic, gupta2017fission}. 
In particular, the positions, widths, and relative strengths of the yield peaks encode the competition among different fission modes~\cite{mcglynn2025extraction}.
Fission modes therefore provide a useful framework for interpreting these multimodal distributions, especially in actinides, where asymmetric yields are commonly associated with competing modes~\cite{brosa1990nuclear, schmidt2000relativistic, andreyev2018nuclear}.
Microscopically, different fission modes are generally understood to reflect shell effects acting through the evolving PES. These effects generate multiple valleys on the PES, each associated with a characteristic prescission shape specified by its elongation and asymmetry. During its descent to scission, the nucleus encounters bifurcation points that determine which channel is followed, while the PES is further modified by additional shell stabilization in the nascent fragments~\cite{ichikawa2012contrasting, Bernard2024HgFm}.

Early interpretations attributed enhanced stability to fragment shell closures at classical magic numbers~\cite{brack1972funny, Strutinsky1967_shell}. This picture was used to explain the abrupt transition from asymmetric to symmetric fission in the Fm isotopes through symmetric splits into fragments with $Z=50$ and neutron numbers close to $N=82$~\cite{hoffman1980fragment,hulet1986bimodal}.
However, subsequent studies showed that closed shell fragment effects alone cannot account for observed fission patterns, as highlighted by the unexpected asymmetric fission of $^{180}$Hg~\cite{andreyev2010new,mcdonnell2014excitation,ichikawa2012contrasting}.
Deformed shell effects were then shown to be crucial for asymmetric fission, with octupole deformed fragments around $Z=52\text{--}56$ and $N=88$ stabilizing the asymmetric channel in actinides~\cite{naturescamps2018impact,scamps2019effect}. 
Further studies in the sublead region point to the role of a proton configuration around $Z=36$ in light fragments as governing asymmetric fission dynamics~\cite{kozulin2022fission, morfouace2025asymmetric}. 
More generally, fragment proton shell effects provide an organizing principle for asymmetric fission across pre-actinide and actinide nuclei~\cite{schmitt2021experimental, mahata2022evidence}. 
Theoretical studies further indicated that shell effects act along the fission path and can induce asymmetric shapes before the final fragments are fully formed~\cite{bernard2023hartree, Bernard2024HgFm}. 
The respective roles of shell effects in shaping competing fission channels and selecting the final yield peaks, however, remain to be clarified.

Various theoretical approaches have been developed to describe SF. 
The early semi-empirical models related fragment yields to the statistical competition among configurations near scission~\cite{wilkins1976scission}, and later phenomenological descriptions interpreted them in terms of competing fission modes, often combined with random neck rupture~\cite{brosa1990nuclear}. 
The GEF framework systematized this picture through a unified treatment of fission channels, shell effects, and statistical partitioning~\cite{GEFschmidt2016general}, while macroscopic-microscopic approaches described yield systematics by combining multidimensional PES with stochastic shape evolution~\cite{randrup2011brownian,mumpower2020primary}. 
Nuclear density functional theory (DFT) provides a more microscopic and globally applicable framework, and time-dependent approaches within or closely related to DFT, such as TDGCM+GOA and TDHFB-based evolution, establish a more direct connection between collective dynamics, fragment formation, and final observables~\cite{simenel2014formation,li2025microscopic,tong2022microscopic,verriere2020time}.

In this work, we employ a two-step framework, which couples the microscopic collective dynamics provided by DFT with Langevin evolution, offering a practical route to linking fission fragment yields to the underlying fission mechanism~\cite{sadhukhan2016microscopic, sadhukhan2017formation}.
Within the DFT picture, the first step in the SF evolution of the nuclear system is tunneling through a multidimensional PES within the WKB approximation, governed by the collective inertia, whereas the second step is the dissipative evolution from the outer turning points to scission via Langevin dynamics. 
We apply this framework to the SF of $^{240}$Pu, $^{232}$Th, and $^{264}$Fm to investigate the interplay of PES topology, collective dynamics, fragment yield formation, and fragment shell effects in fission channel competition. 
These nuclei are chosen to represent distinct SF patterns: $^{240}$Pu as a benchmark asymmetric case, $^{232}$Th as a transitional system with broader asymmetric yields, and $^{264}$Fm as a symmetric case associated with strong shell stabilization in the vicinity of $^{132}$Sn~\cite{chen2026impact,fm264washiyama2024spontaneous}. 
Section~\ref{Framework} presents the theoretical framework, Sec.~\ref{Results} discusses the calculated PES, fragment yields, and shell analyses for $^{240}$Pu, $^{232}$Th, and $^{264}$Fm, and Sec.~\ref{Conclusion} summarizes the main results and provides a brief outlook.

\section{Framework}
\label{Framework}

\subsection{HFB calculations and PES}
\label{subsec:hfb_pes}

The PES is obtained from constrained HFB calculations performed with the axially symmetric solver HFBTHO~\cite{HFBTHO4.0marevic2022axially}.
We employ the SkM$^{\ast}$ parametrization of the Skyrme EDF together with a density-dependent mixed pairing interaction, with pairing correlations treated using the Lipkin--Nogami prescription~\cite{pes240puschunck2014description,stoitsov2003systematic}.
The collective space is spanned by the axial quadrupole and octupole moments, $\bm{q}\equiv(Q_{20}, Q_{30})$, and the PES is mapped on a uniform $(Q_{20}, Q_{30})$ mesh with steps of $1~\mathrm{b}$ and $1~\mathrm{b}^{3/2}$.
The HFB equations are solved in a deformed harmonic-oscillator basis truncated at $N_{\mathrm{sh}}=20$, which is sufficient to ensure numerical convergence of the HFB energies for the present calculations~\cite{perez2017axially,stoitsov2013axially}.
The vibrational zero-point-energy correction is evaluated within the adiabatic time-dependent HFB (ATDHFB) framework. 
The collective potential $V(\bm q)$ is then obtained from the total HFB energy by subtracting this correction and shifting the resulting surface such that the ground-state energy is set to zero~\cite{STASZCZAK1989589,dobaczewski2002nuclear}.

At each mesh point, the constrained HFB solution is obtained by minimizing the expectation value of the constrained Hamiltonian: 
\begin{equation}
	\hat{H}'=\hat{H}_{\mathrm{HFB}}
	-\sum_{\tau=n,p}\lambda_{\tau}\hat{N}_{\tau}
	-\sum_{\mu=2,3}\lambda_{\mu}\hat{Q}_{\mu 0},
	\label{eq:Hamiltonian}
\end{equation}
where $\tau=n,p$ denotes neutrons and protons, $\hat{N}_{\tau}$ is the corresponding particle number operator, $\lambda_{\tau}$ are the particle number Lagrange multipliers, and $\lambda_{\mu}$ are the Lagrange multipliers associated with the multipole constraints.
With $r_\perp^2=x^2+y^2$, the axial multipole moments are defined from the nucleon density $\rho(\bm r)$ as
\begin{equation}
	\begin{aligned}
		Q_{20} &= \int \mathrm{d}^{3}r\, \rho(\bm r)\left(z^{2}-\frac{1}{2}r_\perp^{2}\right),\\
		Q_{30} &= \int \mathrm{d}^{3}r\, \rho(\bm r)\left(z^{3}-\frac{3}{2}z\,r_\perp^{2}\right).
	\end{aligned}
	\label{eq:Q20Q30_def}
\end{equation}

\subsection{WKB tunneling}
\label{subsec:wkb_tunneling}
In the first step, the fission system reaches the outer turning point from the inner turning point by tunneling through the fission barrier. 
We describe this step within the semi-classical WKB approximation, in which the tunneling probability is determined by the collective potential and inertia~\cite{sadhukhan2013spontaneous}.

On the PES, the tunneling trajectory is determined by minimizing the collective action. 
This trajectory defines the least-action path, denoted by $L$. 
We parameterize $L$ in the collective space $\bm q$ by a path coordinate $s$, and obtain it using the dynamic programming method~\cite{baran1981dynamic,flynn2022nudged}.
The action integral is given by~\cite{PhysRevC.22.1979, sadhukhan2013spontaneous}
\begin{equation}
	S(L)=\frac{1}{\hbar}\int_{s_{\mathrm{in}}}^{s_{\mathrm{out}}}
	\sqrt{2\,\mathcal{M}_{\mathrm{eff}}(s)\,\bigl[V(\bm q(s))-E_{0}\bigr]}\,\mathrm{d}s,
	\label{eq:wkb_action}
\end{equation}
where $E_{0}$ is the collective ground-state energy entering the turning-point condition. 
The inner turning point $s_{\mathrm{in}}$ is defined by the condition $V(\bm q)=E_{0}$ near the ground-state minimum, whereas the outer turning points $s_{\mathrm{out}}$ are sampled at equal intervals along the same condition contour at the entrance to the classically allowed region.
In the present work, $E_{0}$ is taken from Refs.~\cite{fm264E0staszczak2009microscopic,fm264washiyama2024spontaneous,sadhukhan2016microscopic}.
The collective inertia tensor $\mathcal{M}_{ij}$ is calculated within the ATDHFB framework using the nonperturbative cranking prescription~\cite{yuldashbaeva1999mass,baran2011quadrupole}. 
The effective inertia $\mathcal{M}_{\mathrm{eff}}(s)$ is then obtained by projecting the collective inertia tensor onto the local tangent to the path:
\begin{equation}
	\mathcal{M}_{\mathrm{eff}}(s)=\sum_{i,j}\mathcal{M}_{ij}\bigl(\bm q(s)\bigr)\,
	\frac{\mathrm{d}q_i}{\mathrm{d}s}\,
	\frac{\mathrm{d}q_j}{\mathrm{d}s}.
	\label{eq:effective_inertia}
\end{equation}
For each $s_{\mathrm{out}}$, the barrier penetration probability is then estimated as $P=\left[1+\exp\bigl(2S(L)\bigr)\right]^{-1}$.
The resulting penetration probabilities provide the statistical weighting for the outer turning point ensemble used to initialize the subsequent Langevin evolution in the classically allowed region.

\subsection{Langevin dynamics and fragments}
\label{subsec:langevin}
In the second step, the fission system evolves in the classically allowed region beyond the outer turning points and descends dissipatively on the PES toward scission.
This stage is described by stochastic Langevin dynamics, which has been widely used to study dissipative fission dynamics and fragment-yield formation, including recent analyses of multimodal fission~\cite{gu2025multimodality, PhysRevC.109.034609}.
In terms of the collective coordinates $q_i$ and their conjugate momenta $p_i$, the coupled Langevin equations read
~\cite{abe1996stochastic,frobrich1998langevin,aritomo2022fission}
\begin{equation}
	\begin{aligned}
		\frac{dq_i}{dt}=&\left(\mathcal{M}^{-1}\right)_{ij}p_j,\\
		\frac{dp_i}{dt}=&-\frac{p_j p_k}{2}\,\frac{\partial}{\partial q_i}\left(\mathcal{M}^{-1}\right)_{jk}
		-\frac{\partial V}{\partial q_i}\\
		&-\eta_{ij}\left(\mathcal{M}^{-1}\right)_{jk}p_k
		+g_{ij}\Gamma_j(t),\\[4pt]
	\end{aligned}
	\label{eq:langevin}
\end{equation}
where $V$ is the collective potential introduced in Sec.~\ref{subsec:hfb_pes}, $\mathcal{M}_{ij}$ is the collective inertia tensor introduced in Sec.~\ref{subsec:wkb_tunneling}, $\eta_{ij}$ is the collective dissipation tensor, and $g_{ij}$ is the random-force strength tensor. 
All these quantities are functions of the collective coordinates $\bm q$.
The time-dependent stochastic variables $\Gamma_i(t)$ are taken as independent Gaussian white noises
\begin{equation}
	\langle \Gamma_i(t)\rangle = 0,\qquad
	\langle \Gamma_i(t)\Gamma_j(t')\rangle=2 \delta_{ij}\,\delta(t-t').
	\label{eq:noise_corr}
\end{equation}
Their strength is fixed by the fluctuation--dissipation relation 
$\sum_k g_{ik}\,g_{jk} = \eta_{ij}k_{B}\,T\,$. 
The collective dissipation tensor $\eta_{ij}$ is taken as a constant tensor, following the prescription in Ref.~\cite{sadhukhan2016microscopic}.
At each time step, the effective temperature $T$ is determined from the intrinsic excitation energy as $k_B T=\sqrt{E^\ast/a}$, with $a=A/10~\mathrm{MeV}^{-1}$ as the level density parameter. 
The intrinsic excitation energy is taken as $E^\ast=V(s_{\mathrm{out}})-V(\bm q)-\frac{1}{2}(\mathcal{M}^{-1})_{ij}p_i p_j$.
The derivatives of the collective potential and the inverse inertia tensor with respect to the collective coordinates, as required in Eq.~(\ref{eq:langevin}), are evaluated by the finite difference method on the same collective mesh.

Langevin trajectories are initiated from an ensemble of outer turning points that mark the entrance to the classically allowed region. 
For each starting point, the stochastic force causes trajectories to diverge, thereby generating a distribution of dissipative evolutions.
Once a Langevin trajectory reaches the scission line, its evolution is terminated, and the corresponding point is taken as the scission configuration. 
The scission line is identified through a neck criterion based on the neck operator, which reads
\begin{equation}
	Q_N=\int \mathrm{d}^3r\,\rho(\bm r)\,
	\exp\!\left[-\frac{(z-z_N)^2}{a_N^2}\right],
	\label{eq:neck_operator}
\end{equation}
where $Q_N$ measures the nucleon content in a Gaussian slice centered at the thinnest part of the neck, $z=z_N$.
We adopt $a_N=1~\mathrm{fm}$ and define scission by the condition $Q_N=2$~\cite{scissionhan2021scission}, which corresponds to a dilute neck close to scission.
Previous Langevin calculations showed that the yield distributions are relatively stable with respect to this criterion~\cite{sadhukhan2016microscopic}.

For each scission configuration, the density is partitioned at the neck plane into two spatial regions, $\Omega_f$ with $f\in\{\mathrm{HF}, \mathrm{LF}\}$, corresponding to the heavy fragment ($\mathrm{HF}$) and its complementary light fragment ($\mathrm{LF}$).
The fragment mass and charge numbers are obtained by integrating the nucleon and proton densities over $\Omega_f$.
Fragment deformations are characterized by first computing the center of mass $z_f$ and then evaluating the intrinsic moments $Q_{20}^{(f)}$ and $Q_{30}^{(f)}$ from Eq.~(\ref{eq:Q20Q30_def}), with the integration restricted to $\Omega_f$, the longitudinal coordinate shifted to $z-z_f$, and $r_\perp$ unchanged~\cite{perez2017axially}.
Fragment yields are constructed by histogramming the resulting scission configurations, normalizing the distributions, and folding the raw yields with a Gaussian of width $\sigma$ to account for particle number fluctuations in the neck at scission, with $\sigma_A=3$ for mass yields and $\sigma_Z=2$ for charge yields~\cite{sadhukhan2016microscopic}.

\subsection{Smoothed level density shell indicator}

\label{subsec:sld}
Single-particle levels in the vicinity of the Fermi energy provide a sensitive probe of shell effects and their evolution with deformation~\cite{Strutinsky1967_shell, xu2024coexistence,naturescamps2018impact}.
The smoothed level density (SLD) shell indicator $\eta_\tau$ ($\tau=n,p$) is therefore employed to quantify the shell structure associated with fragment configurations at scission~\cite{bernard2023hartree, Bernard2024HgFm}. 

\begin{figure}[t]
	\centering    \includegraphics[width=1\singlefigwidth]{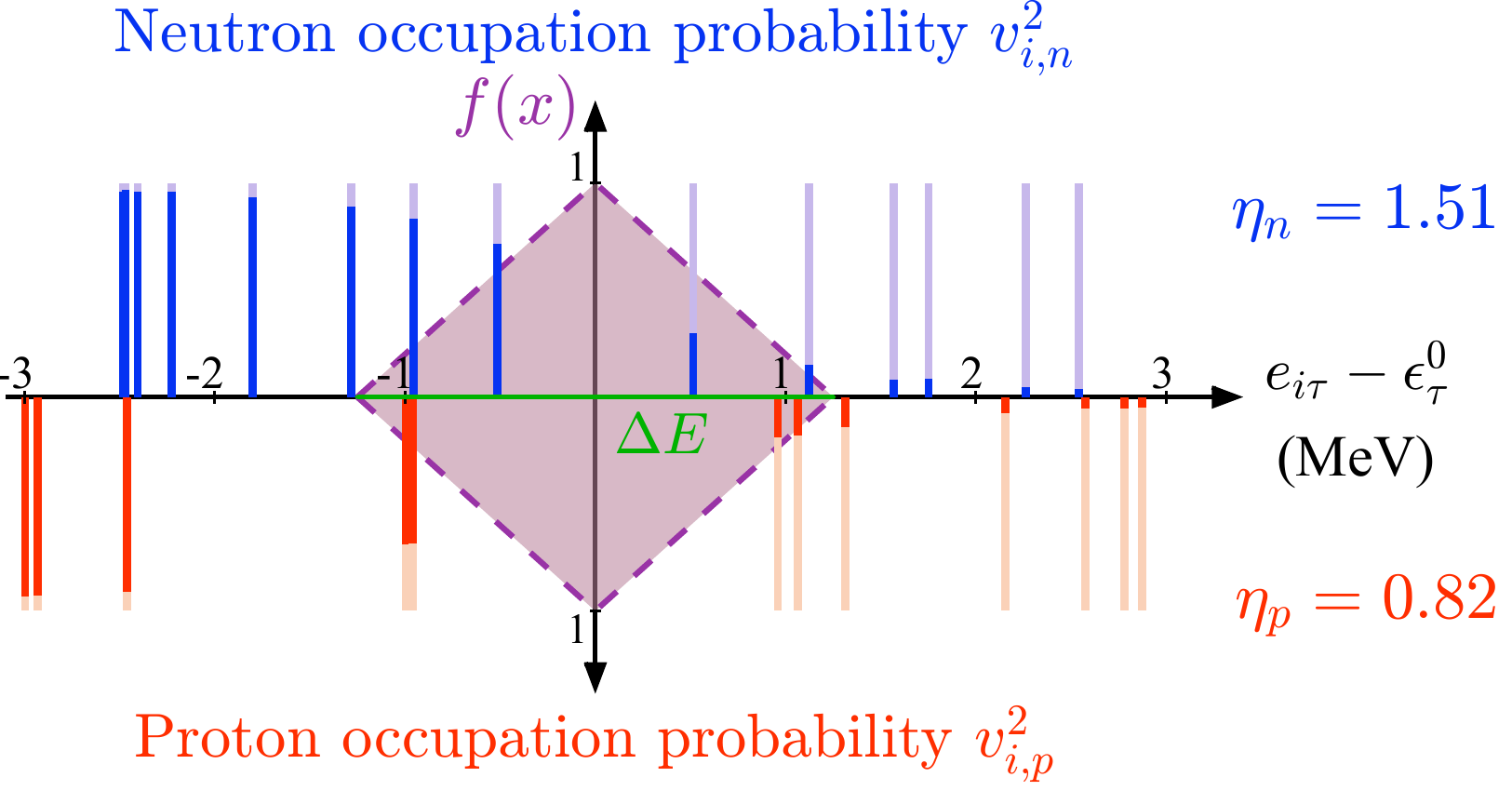}
\caption{
    Schematic construction of the SLD indicator for $^{144}$Ba.
	The horizontal axis shows the single-particle energies relative to the reference energy, $e_{i\tau}-\epsilon_\tau^0$.
    The triangular smoothing function $f(x)$ is represented by the light-purple triangular region, with width $\Delta E$ indicated by the green line.
    The occupation probabilities $v_{i,\tau}^2$ are indicated by the opaque parts of the vertical lines.
	The resulting $\eta_n$ and $\eta_p$ are shown on the right.
}
	\label{fig:sld_indicator}
\end{figure}

A reference energy $\epsilon_\tau^{0}$ is taken as the midpoint between the two single-particle levels adjacent to the Fermi energy.
The SLD indicator is then written as
\begin{equation}
	\eta_\tau=\sum_i f\!\left(e_{i\tau}-\epsilon_\tau^{0}\right),
	\label{eq:sld_eta}
\end{equation}
where $e_{i\tau}$ denotes the single-particle energies inside the chosen energy window $\Delta E$.
A triangular smoothing function is introduced, 
\begin{equation}
	f(x)=
	\begin{cases}
		1-|x|/E_w, & |x|\le E_w,\\
		0, & |x|>E_w,
	\end{cases}
	\qquad E_w=\Delta E/2.
	\label{eq:sld_weight}
\end{equation}
The resulting $\eta_\tau$ is a dimensionless weighted count of the single-particle levels within the energy window and varies smoothly when levels move across the window boundaries.
The window width is taken as $\Delta E=2.5~\mathrm{MeV}$, following Ref.~\cite{Bernard2024HgFm}, which is of the order of typical shell gaps. 
A smaller $\eta_\tau$ corresponds to fewer single-particle levels contributing within the smoothing window, and therefore to a lower level density near the Fermi surface and a stronger shell effect, whereas a larger $\eta_\tau$ indicates weaker shell structure. 
This construction is illustrated schematically in Fig.~\ref{fig:sld_indicator} for $^{144}$Ba at $Q_{20}=12~\mathrm{b}$ and $Q_{30}=2.5~\mathrm{b}^{3/2}$, selected from a probability region of the Langevin scission configuration deformation distribution in the $^{240}$Pu calculation.
In this example, the relatively large gap around $\epsilon_p^0$ corresponds to fewer proton levels within the smoothing window, resulting in a smaller $\eta_p$ and reflecting a stronger proton shell effect at this deformation.

The SLD indicators are evaluated for the individual heavy and light fragments at scission rather than for the compound system.
Fragments identified from the Langevin scission configurations are grouped according to their particle numbers, and constrained HFB calculations are performed for the corresponding isolated fragment species using the same Skyrme EDF, pairing, and basis settings as in Sec.~\ref{subsec:hfb_pes}, with quadrupole and octupole constraints.
The resulting HFB solutions define the fragment PES, and their single-particle levels are used to construct the corresponding $\eta_{\tau}^{(f)}$ maps on the $(Q_{20},Q_{30})$ plane.
The Langevin scission configurations are then projected onto corresponding PES and SLD maps to identify shell-stabilized scission configurations associated with the observed yield peaks.

\section{Results}
\label{Results}

\begin{figure}[htbp]
\centering
\includegraphics[width=1\singlefigwidth]{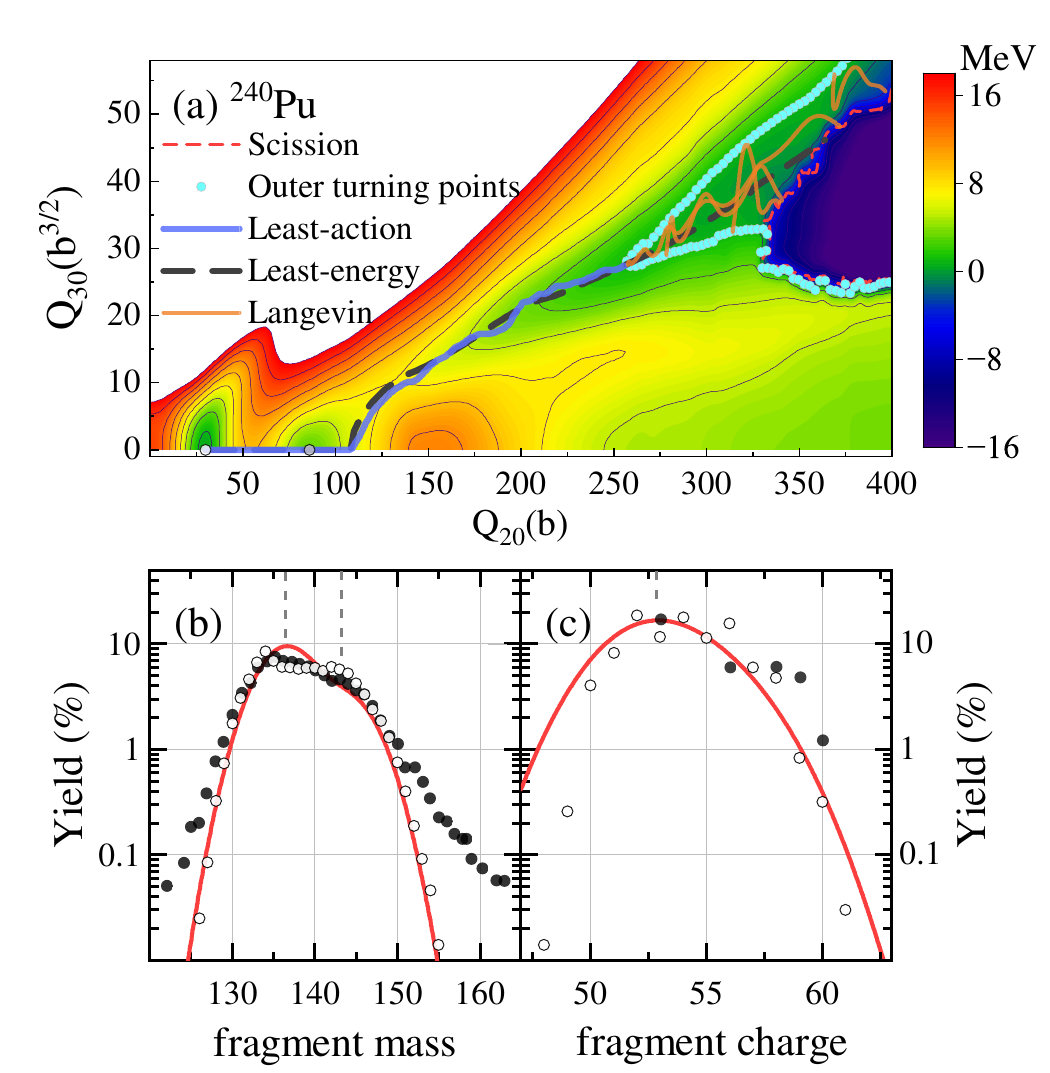}
\caption{PES and heavy fragment yields for the SF of $^{240}$Pu.
Panel (a) shows the PES in the collective $(Q_{20},Q_{30})$ plane.
Colors (MeV) and contour lines are given relative to the ground state.
  The white dot marks the inner turning point, the gray dot marks the fission isomer, the cyan dots mark the outer turning points, the blue curve marks the least-action path, the black dashed curve marks the least-energy path along the line of local minima on the PES, the orange curves show the Langevin trajectories, and the red dashed line marks the scission line.
  Panels (b) and (c) show the heavy fragment mass and charge yields.
  Red solid curves denote results of the present calculation, open circles denote the GEF predictions (2025/V1.2), and filled circles denote the experimental data from Refs.~\cite{thierens1981kinetic,laidler1962mass}. All yields are plotted on a logarithmic scale.
  }
  \label{fig:Pu240_overview}

\end{figure}

\begin{figure}[htbp]
  \centering    \includegraphics[width=1\singlefigwidth]{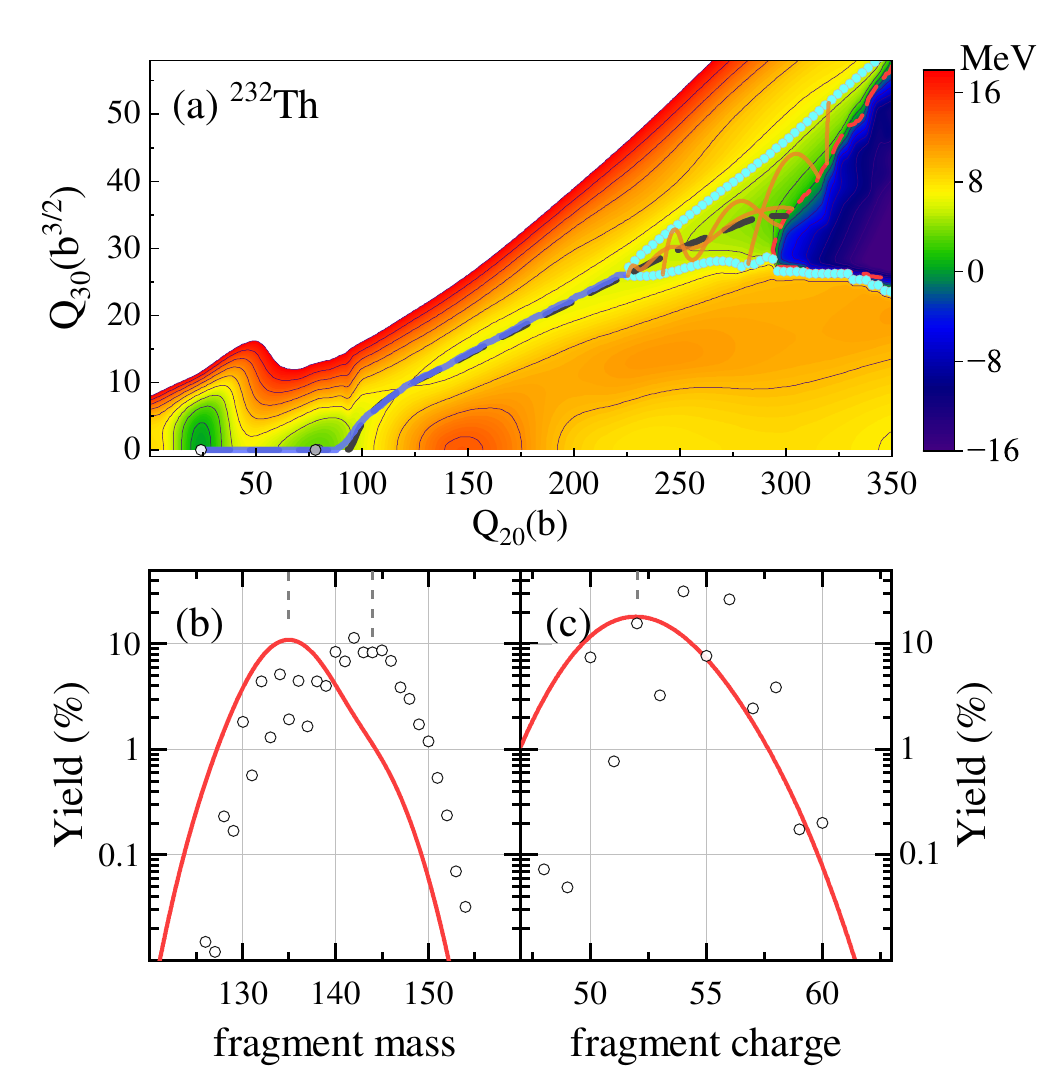}
  \caption{Same as Fig.~\ref{fig:Pu240_overview}, but for $^{232}$Th.
  Panels (b) and (c) show the heavy fragment mass and charge yields.
  Only the results of the present calculation and the GEF predictions are shown.
}
 
  \label{fig:Th232_overview}
\end{figure}

\begin{figure}[htbp]
  \centering    \includegraphics[width=1\singlefigwidth]{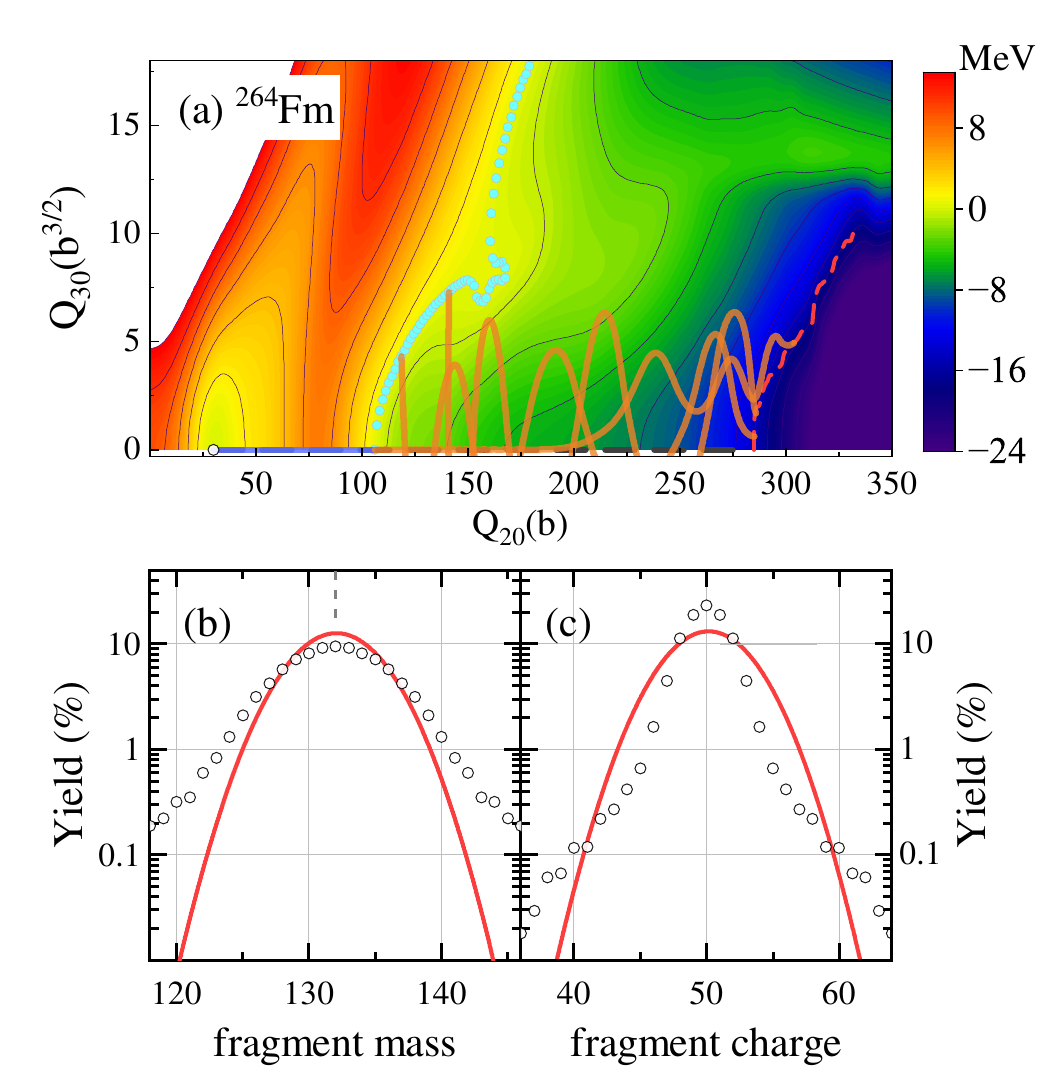}
  \caption{Same as Fig.~\ref{fig:Pu240_overview}, but for $^{264}$Fm.
  Panels (b) and (c) show the full mass and charge yields.
  Only the results of the present calculation and the GEF predictions are shown.}
  \label{fig:Fm264_overview}
\end{figure}

\subsection{PES and Fragment Yields}

In this section, we discuss the connection between PES topology, Langevin dynamics, and fragment yields within the present two-step framework.
Figures~\ref{fig:Pu240_overview}--\ref{fig:Fm264_overview} display the PES and the corresponding fragment yields for $^{240}$Pu, $^{232}$Th, and $^{264}$Fm.
The benchmark case $^{240}$Pu is first used to illustrate the connection among shell effects, fission valleys, trajectory populations, and the final yield distribution.
The same discussion is then extended to $^{232}$Th and $^{264}$Fm.

\subsubsection{$^{240}$Pu}
\label{subsubsec:Pu240_yields}
The SF of $^{240}$Pu serves as a benchmark case for the present framework. 
As shown in Figure~\ref{fig:Pu240_overview}(a), the PES exhibits a prolate minimum near $(Q_{20}, Q_{30})\approx(25,0)$, at which the inner turning point is located, and a secondary minimum near $(80,0)$ corresponding to the fission isomer. 
Previous multidimensional studies of $^{240}$Pu have shown that, between the inner turning point and the fission isomer, the least-action path remains essentially axial, whereas the least-energy path passes through triaxial shapes near the first barrier~\cite{sadhukhan2016microscopic}.
In the present collective $(Q_{20}, Q_{30})$ plane, both paths are projected onto the same axial plane and therefore appear nearly overlapping before the fission isomer.
This difference may influence the calculated SF half-lives; here, we focus on the subsequent Langevin evolution, which determines the fission channel and the scission configurations used for the fragment yields and shell analysis.

The deviation from the symmetric direction starts around $(110,0)$, as the barrier near $(150,0)$ hinders the symmetric route and favors evolution toward larger $Q_{30}$.
Once the system enters the asymmetric valley, the least-action path continues along this valley, while the ridge extending approximately from $(150,0)$ toward $(400,20)$ prevents a return to the symmetric region.
This ridge-valley structure reflects the competition between symmetric and asymmetric pathways on the evolving PES and is consistent with shell effects acting through the PES to drive the system toward asymmetric shapes~\cite{Bernard2024HgFm}.

Starting from the outer turning points, Langevin trajectories describe the descent to scission and connect the dynamics in the classically allowed region with the final fragment yields.
To obtain the fission yields, 80\,000 Langevin trajectories were calculated from 200 outer turning points sampled at equal intervals.
The four representative Langevin trajectories shown as orange lines in Fig.~\ref{fig:Pu240_overview}(a) illustrate that the trajectory ensemble primarily populates two regions, around $Q_{30}\approx35$ and $Q_{30}\approx50$, indicating two favored asymmetric fission channels.
The resulting scission configurations were used to construct the heavy fragment mass and charge yields shown in Figure~\ref{fig:Pu240_overview}(b) and (c).

The fragment mass yield in the present calculation, shown by the red solid curve, is dominated by a peak at $A_H\approx134\text{--}136$, while retaining appreciable strength up to $A_H\approx140\text{--}145$, indicating a dominant asymmetric component and a weaker subdominant component on the heavier mass side.
These two mass components can be traced back to the two principal trajectory populations on the PES. The lower $Q_{30}$ region captures the larger fraction of trajectories and feeds the dominant component, whereas the higher $Q_{30}$ region is less populated and contributes to the weaker shoulder on the heavier mass side.
The charge yield does not resolve the two components as distinctly, but the main strength is concentrated at $Z_H\approx52\text{--}56$, with a weaker extension toward heavier charge.

In the present calculation, the dominant and subdominant heavy fragment mass components of $^{240}$Pu lie in the mass regions associated with the semi-classical Brosa picture with the standard~I (S1) mode, centered at $A_H\approx134\text{--}136$, and the standard~II (S2) mode, centered at $A_H\approx140\text{--}141$~\cite{brosa1990nuclear,dematte1997fragments,wilkins1976scission,  mahata2022evidence}.
In this phenomenological picture, the standard fission family in actinides commonly includes the asymmetric S1 and S2 modes within a single nucleus, which are typically identified through the combined systematics of mass yields, total kinetic energy, and mode decompositions. 
The S1 mode is generally associated with a less asymmetric split and a more compact scission configuration. It is often linked to the spherical neutron shell closure at $N=82$. In contrast, the S2 mode is usually associated with a more asymmetric and more elongated scission shape and is commonly related to deformed shell stabilization around $N=88$.
The present calculation does not attempt a full Brosa mode reconstruction; instead, this phenomenological picture is used only as a point of reference for the discussion below.

The GEF predictions, shown as open circles, are used here as a phenomenological reference for fission yields, since GEF incorporates empirical fission systematics with parameters fitted to extensive experimental data and has been widely applied to describe yield distributions.
The present result, the GEF prediction, and the experimental data agree well on the peak positions and relative strengths of the two mass components.
The main discrepancy appears in the tails: the experimental distribution is broader on both sides, whereas both calculations remain more localized around the main mass components.

\subsubsection{$^{232}$Th}
\label{subsubsec:Th232_yields}

$^{232}$Th also favors asymmetric fission, with a PES topology qualitatively similar to that of $^{240}$Pu. 
As shown in Fig.~\ref{fig:Th232_overview}(a), the barrier near $(150,0)$ and ridge extending toward the large deformation region around $(350,20)$ hinder the symmetric route and steer the evolution into an asymmetric valley.

Fig.~\ref{fig:Th232_overview}(b) shows that the heavy fragment mass yield in the present calculation is dominated by a broad peak at $A_H\approx134\text{--}136$.
Toward increasing $A_H$, the yield decreases, with a slightly weaker slope near $A_H\approx144$, indicating a weak heavier mass shoulder rather than a clearly separated second peak.
This mass yield pattern can be interpreted phenomenologically in terms of two mass regions commonly associated in the Brosa picture with S1 mode at $A_H\approx135$ and S2 mode at $A_H\approx139\text{--}143$.
The charge yield in Fig.~\ref{fig:Th232_overview}(c) is concentrated at $Z_H\approx52\text{--}54$ and shows a tail toward heavier charge.
The GEF prediction supports a two-component interpretation with similar mass peak positions, but gives more comparable component strengths, whereas the present calculation places more weight on the lower mass side.
Correspondingly, the present charge yield is more concentrated on the lower charge side.
The difference mainly reflects how the relative component weights are obtained: GEF prediction assigns them phenomenologically, whereas the present calculation obtains them from the PES topology and the Langevin trajectory population.
In the present calculation, the PES gradient from the outer turning points toward the scission line is smaller than in $^{240}$Pu. 
The Langevin trajectories are therefore concentrated mainly in the lower $Q_{30}$ region of the asymmetric valley, which feeds the dominant lower mass component, whereas the nearby higher $Q_{30}$ region is also populated but appears only as a shoulder on the heavier mass side rather than as a clearly separated peak.

A previous static microscopic fission study based on the Skyrme EDF and nuclear localization function analysis showed that a hyperdeformed configuration of $^{232}$Th along the fission path may already exhibit a quasimolecular arrangement resembling the spherical shell prefragments $^{132}$Sn and $^{100}$Zr well before scission~\cite{zhang2016nucleon}. 
In the present dynamical calculation, however, the final yields are built from an ensemble of Langevin scission configurations, and neither the present yields nor the GEF prediction shows a pronounced final peak at such a split. 
This contrast suggests that an early shell-favored prefragment along the static fission path may not, by itself, be sufficient to determine the final yield maximum. 
Instead, the final yield pattern appears to depend on which scission configurations remain accessible and are populated during the dynamic evolution through the classically allowed region.

\subsubsection{$^{264}$Fm}
\label{subsubsec:Fm264_yields}

In contrast to $^{232}$Th, $^{264}$Fm provides a clear symmetric fission case. 
As shown in Fig.~\ref{fig:Fm264_overview}(a), the least-action path enters the symmetric valley and then proceeds toward scission primarily with increasing $Q_{20}$ while remaining close to $Q_{30}\approx0$.  %along $Q_{20}\approx0$.
The classically allowed region develops a pronounced symmetric valley, without a competing asymmetric valley of comparable depth. 
The Langevin trajectories are confined to a narrow region around the symmetric path, thereby favoring symmetric fission.
Figures~\ref{fig:Fm264_overview}(b) and (c) show narrow symmetric mass and charge yields centered at $A\approx132$ and $Z\approx50$, corresponding to a symmetric split in the vicinity of the $^{132}$Sn configuration. 
The present calculation and GEF prediction locate the maxima at nearly the same mass and charge, but the widths differ: the present mass yield is more concentrated around the symmetric split, and the GEF prediction is more concentrated in charge yield.

Thus, $^{264}$Fm provides a case in which the spherical shell-favored configuration is directly reflected in the final symmetric yield peak, consistent with the static nuclear localization picture of fragment formation~\cite{zhang2016nucleon}.

\subsection{Fragment Shell Analysis}

The final fragment distribution reflects not only the macroscopic collective dynamics of the fissioning system, but also the microscopic shell structure and deformation properties of the fragments~\cite{naturescamps2018impact}.
To elucidate the microscopic origin of the yield components discussed above, a fragment shell analysis is carried out for representative even--even fragment pairs selected from each fission channel. 

The selection is based on the Langevin scission configurations.
For each configuration, the particle numbers of the heavy and light fragments define a fragment split; configurations leading to the same split are then accumulated to obtain their statistical weights in the fragment ensemble.
For a given yield component, the relevant mass region is identified above as the S1 or S2 channel, and the even--even split with the largest statistical weight in that region is taken as the representative pair.
The S1 and S2 channels are represented by $^{134}$Te/$^{106}$Mo and $^{144}$Ba/$^{96}$Sr for $^{240}$Pu, by $^{136}$Te/$^{96}$Sr and $^{140}$Xe/$^{92}$Kr for $^{232}$Th, respectively. 
The dominant symmetric channel of $^{264}$Fm is represented by the $^{132}$Sn/$^{132}$Sn split.

\begin{figure}[htbp]
	\centering
    \includegraphics[width=1\singlefigwidth]{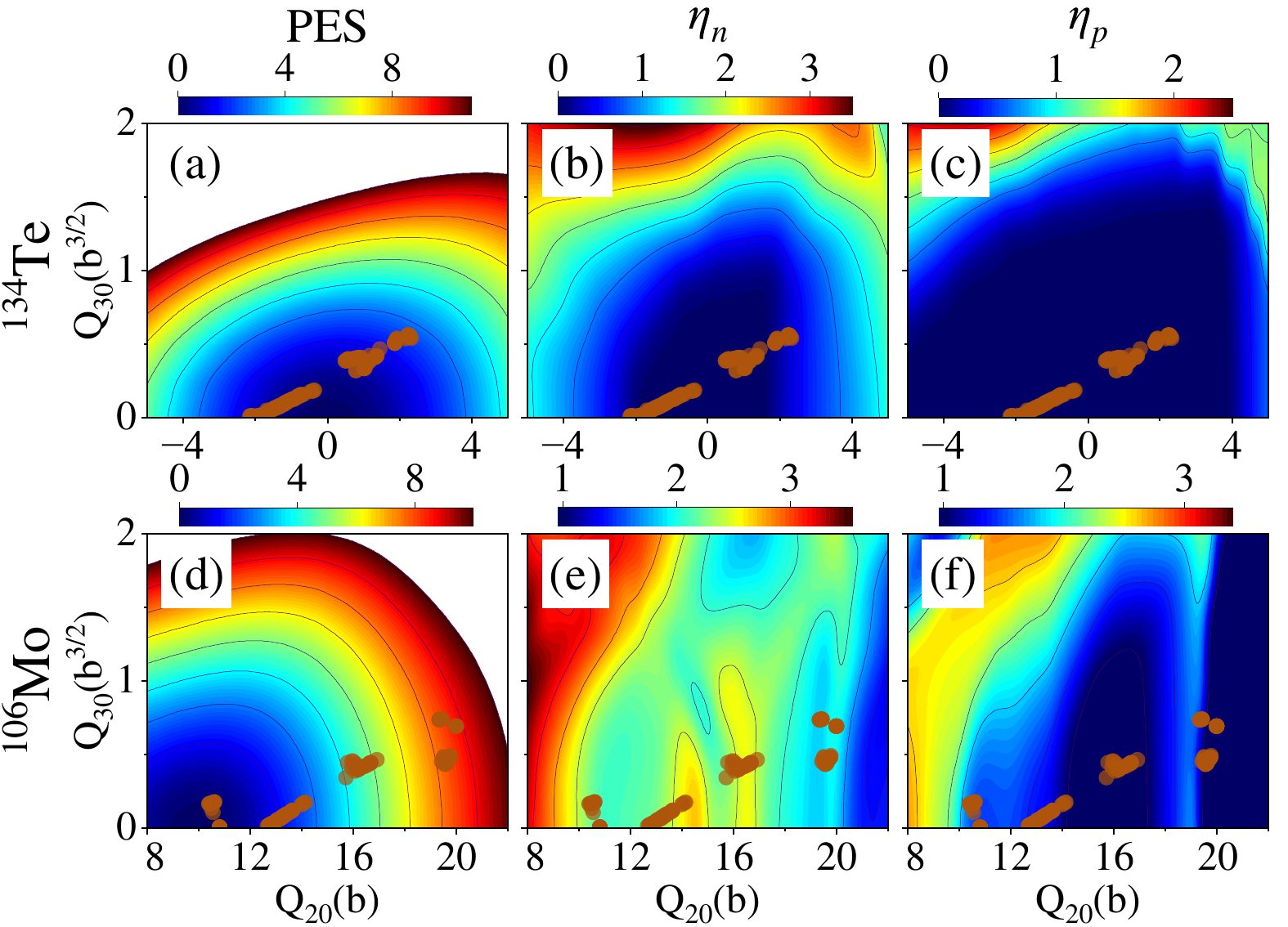}
\caption{Fragment PES, $\eta_n$, and $\eta_p$ maps on the $(Q_{20},Q_{30})$ deformation plane for the fragment pair $^{134}$Te [panels (a)--(c)] and $^{106}$Mo [panels (d)--(f)] in the S1 channel of $^{240}$Pu.
Colors (MeV) and contour lines in the PES panels are given relative to the ground state.
Orange markers denote the deformation of the Langevin scission configurations associated with this channel.
Blue regions correspond to lower values of the energy, $\eta_n$, and $\eta_p$.
The color scales are chosen separately for each panel.}
	\label{fig:Pu240_S1_shell}
\end{figure}

\begin{figure}[htbp]
	\centering
    \includegraphics[width=1\singlefigwidth]{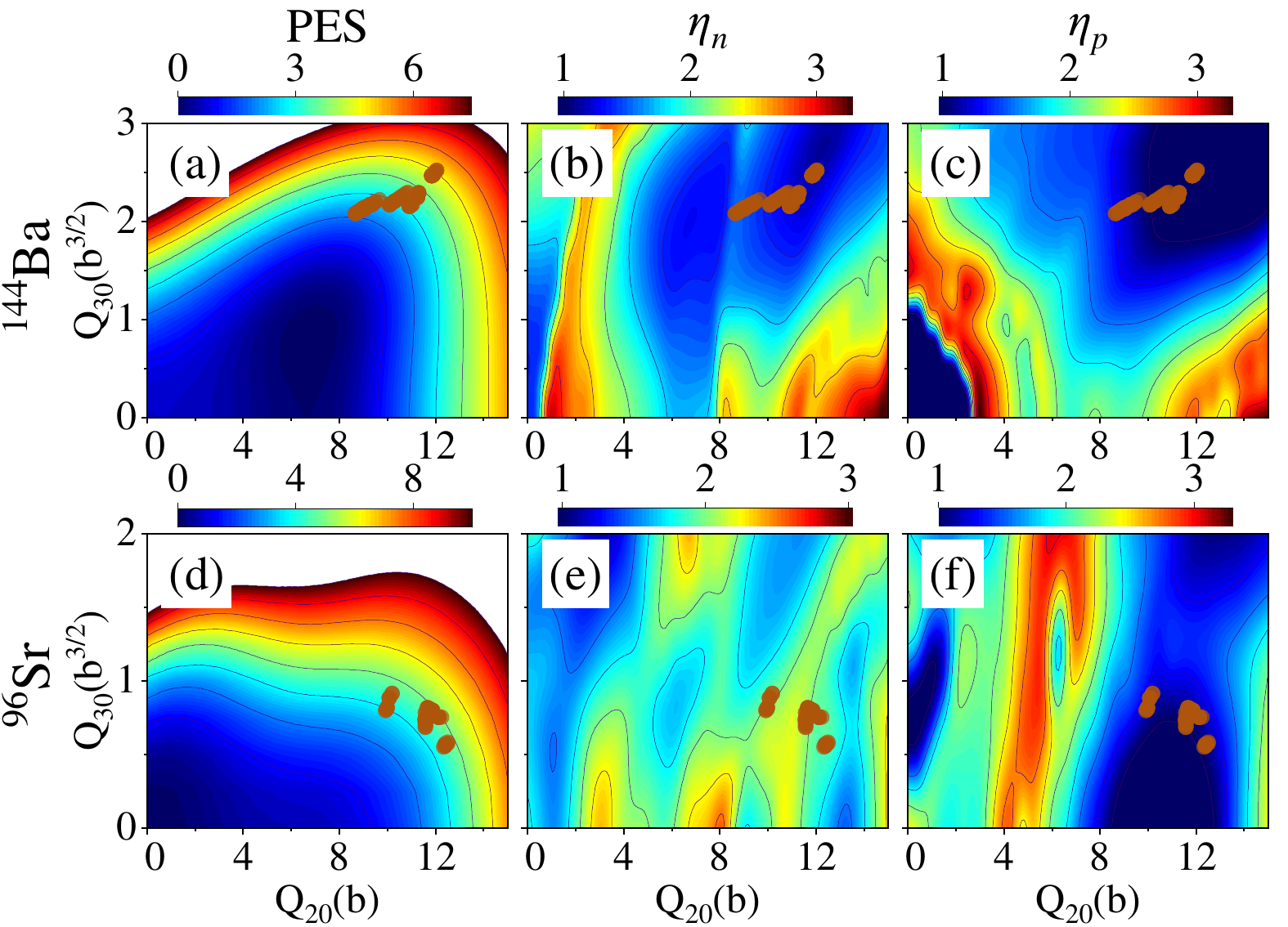}
	\caption{The same as Fig.~\ref{fig:Pu240_S1_shell}, but for fragment pair $^{144}$Ba/$^{96}$Sr in the S2 channel of $^{240}$Pu.}
	\label{fig:Pu240_S2_shell}
\end{figure}

\begin{figure}[htbp]
	\centering
    \includegraphics[width=1\singlefigwidth]{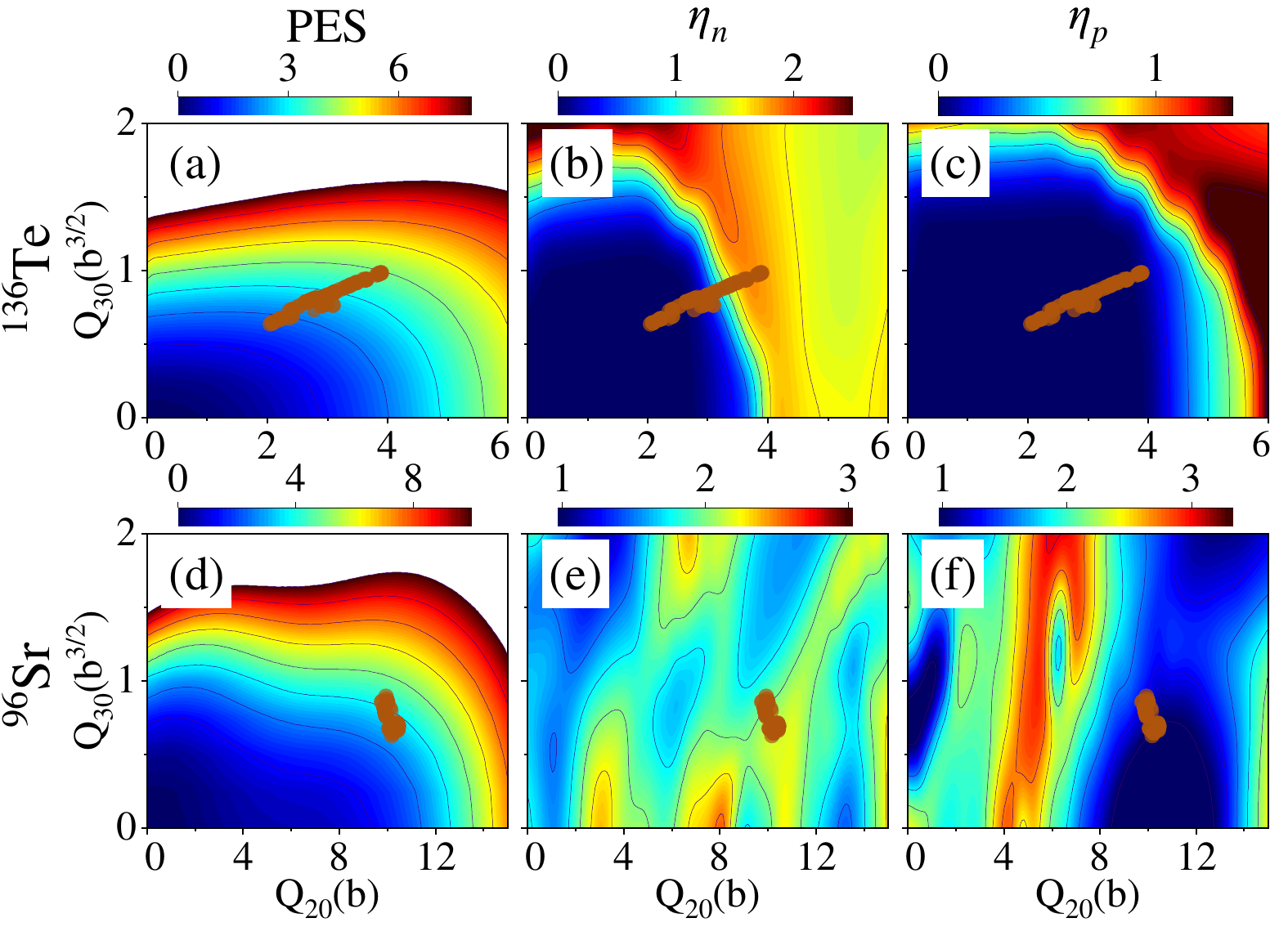}
	\caption{The same as Fig.~\ref{fig:Pu240_S1_shell}, but for fragment pair $^{136}$Te/$^{96}$Sr in the S1 channel of $^{232}$Th.}
	\label{fig:Th232_S1_shell}
\end{figure}

\begin{figure}[htbp]
	\centering
    \includegraphics[width=1\singlefigwidth]{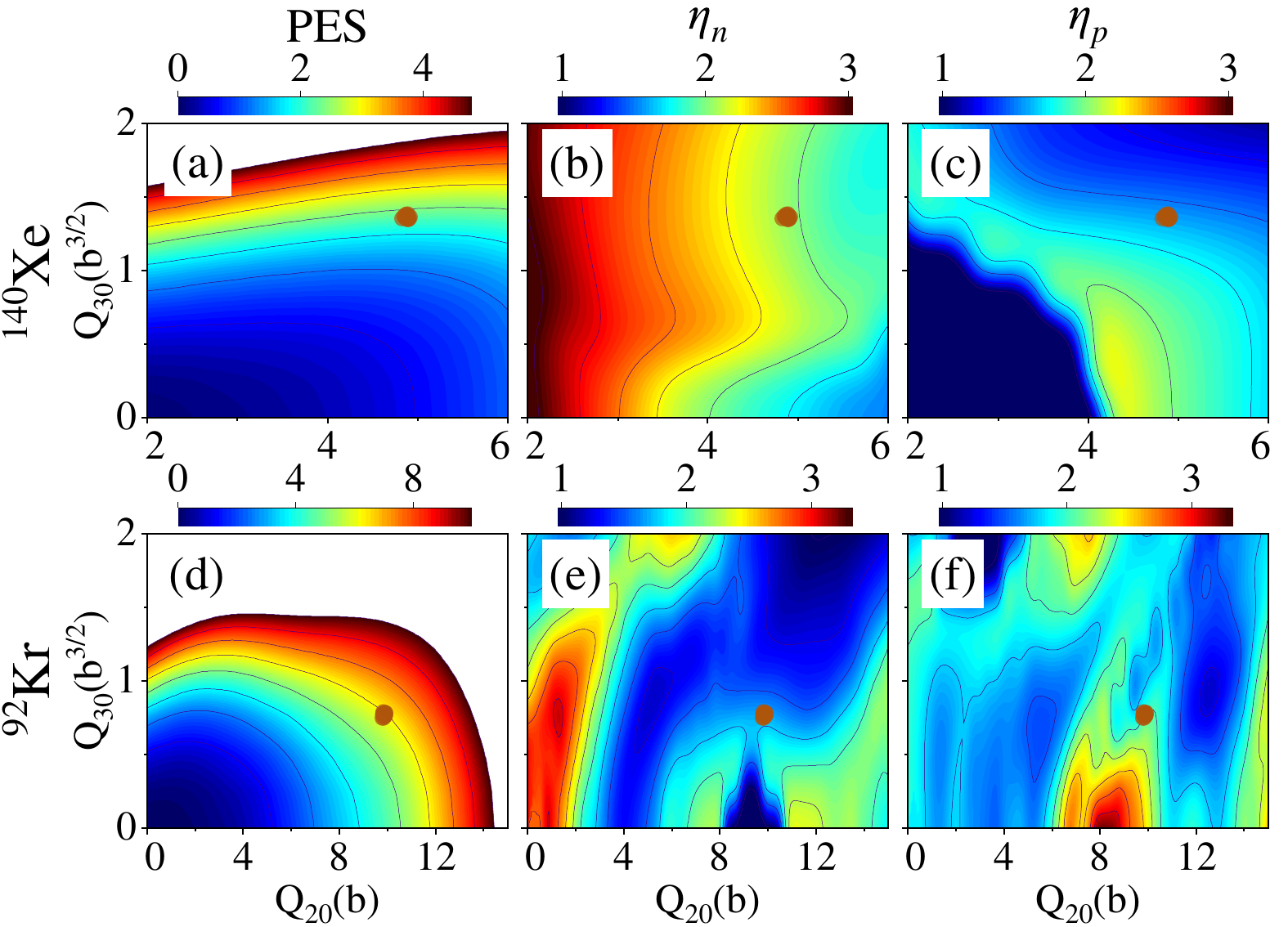}
	\caption{The same as Fig.~\ref{fig:Pu240_S1_shell}, but for fragment pair $^{140}$Xe/$^{92}$Kr in the S2 channel of $^{232}$Th.}
	\label{fig:Th232_S2_shell}
\end{figure}

\begin{figure}[htbp]
	\centering
    \includegraphics[width=1\singlefigwidth]{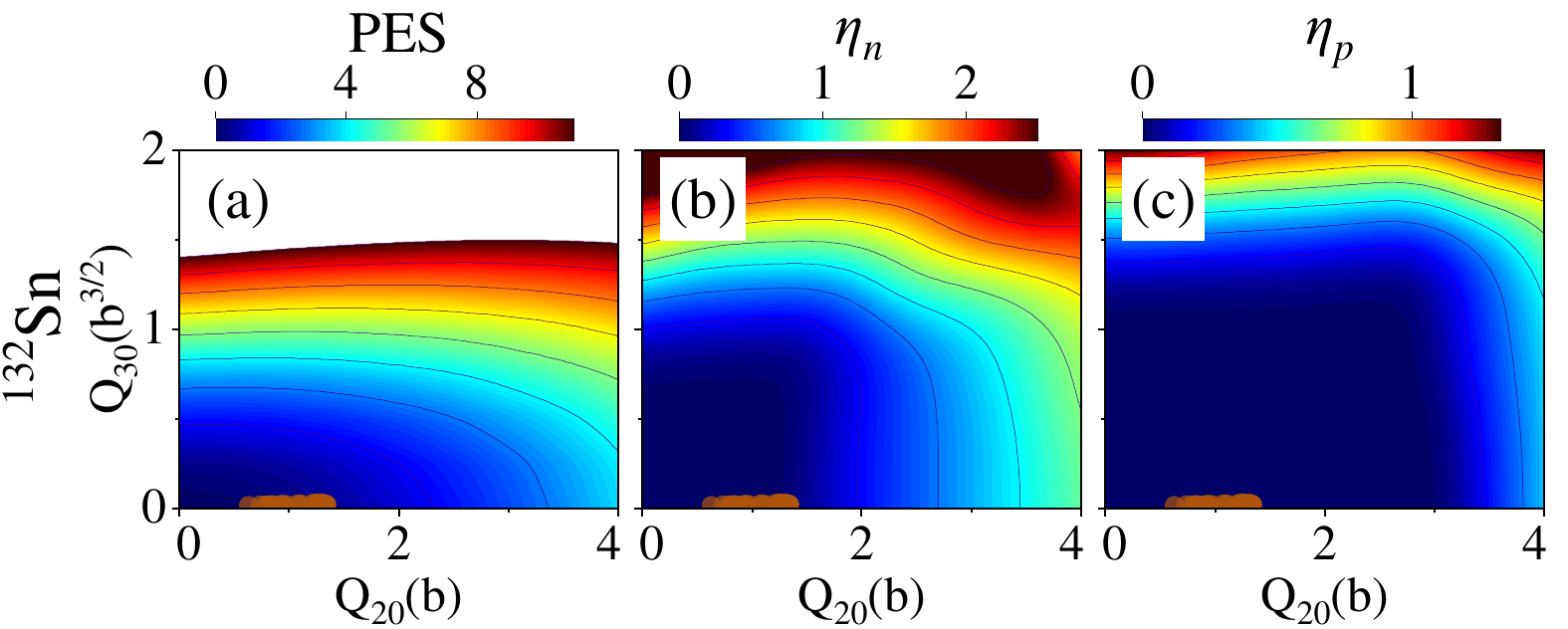}
	\caption{The same as Fig.~\ref{fig:Pu240_S1_shell}, but for fragment pair $^{132}$Sn/$^{132}$Sn in the dominant symmetric channel of $^{264}$Fm.}
	\label{fig:Fm264_sym_shell}
\end{figure}

For each channel, the heavy and light fragments are treated separately. 
Their intrinsic $(Q_{20}, Q_{30})$ values are extracted from the Langevin scission configurations, while constrained HFB calculations for the corresponding isolated fragment nuclei are used to construct the fragment PES, the associated neutron SLD maps $\eta_n(Q_{20}, Q_{30})$, and the proton SLD maps $\eta_p(Q_{20}, Q_{30})$, as introduced in Sec.~\ref{subsec:sld}.
To characterize the dynamically selected fragment configurations on the fragment PES, the deformation energy of each isolated fragment is evaluated. 
For each fragment $f$, the deformation energy is defined as the HFB energy of the fission fragment at the specified Langevin scission deformation minus the corresponding ground-state energy:
\begin{equation}
E_{\mathrm{def}}^{(f)}(Q_{20},Q_{30})
=E_{\mathrm{FF}}^{(f)}(Q_{20},Q_{30})-E_{\mathrm{g.s.}}^{(f)},
\label{eq:fragment_deformation_energy}
\end{equation}
% Then, the average deformation energy is evaluated as
The average deformation energy is then evaluated as
\begin{equation}
\bar{E}_{\mathrm{def}}^{(f)}
=\frac{1}{N_f}\sum_{k=1}^{N_f}
E_{\mathrm{def}}^{(f)}(Q_{20,k}^{(f)},Q_{30,k}^{(f)}),
\label{eq:average_fragment_deformation_energy}
\end{equation}
where $N_f$ is the number of Langevin scission configurations.
The same averaging procedure is used to obtain the average SLD, $\bar{\eta}_{n}^{(f)}$ and $\bar{\eta}_{p}^{(f)}$. 
For direct comparison, the deformations of Langevin scission configurations are then projected onto the PES and SLD maps.
This procedure makes it possible to assess whether the dynamically selected configurations are mainly driven by neutron and/or proton shell stabilization, or their combined action, as indicated by low $\eta_n$ and/or low $\eta_p$ regions.
Figures~\ref{fig:Pu240_S1_shell}--\ref{fig:Fm264_sym_shell} summarize this analysis for the dominant yield channels of $^{240}$Pu, $^{232}$Th, and $^{264}$Fm.

\subsubsection{$^{240}$Pu}

Figure~\ref{fig:Pu240_S1_shell} shows the fragment PES, $\eta_n$, and $\eta_p$ maps for the S1 channel, represented by $^{134}$Te [panels (a)--(c)] and $^{106}$Mo [panels (d)--(f)]. 
For the heavy fragment $^{134}$Te ($Z=52$, $N=82$), the ground state is located at the spherical configuration.
The Langevin scission configurations, shown as orange markers, cluster around the spherical ground state, yielding a small average deformation energy, $\bar{E}_{\mathrm{def}}=0.90$ MeV, together with low average SLD, $\bar{\eta}_n=0.04$ and $\bar{\eta}_p=0.00$.
These consistently small values of $\bar{E}_{\mathrm{def}}$, $\bar{\eta}_n$, and $\bar{\eta}_p$ indicate strong microscopic selectivity for a near-spherical shell-stabilized configuration in which the spherical ground state is reinforced by both the neutron shell closure at $N=82$ and proton shell stabilization around $Z=52$.
For the complementary light fragment $^{106}$Mo ($Z=42$, $N=64$), the fragment ground state is located at $Q_{20}\approx10~\mathrm{b}$ and $Q_{30}=0 ~\mathrm{b}^{3/2}$, corresponding to a prolate configuration.
The scission configurations are confined to small $Q_{30}$ but extend over a broader interval in $Q_{20}$, forming three characteristic deformation groups.
The first group lies near the prolate ground state around $Q_{20}\approx12~\mathrm{b}$ and has a small average deformation energy, $\bar{E}_{\mathrm{def}}^{(1)}=0.94$ MeV, with $\bar{\eta}_n^{(1)}=2.18$ and $\bar{\eta}_p^{(1)}=1.33$.
The second group is centered around $Q_{20}\approx16~\mathrm{b}$, with $\bar{E}_{\mathrm{def}}^{(2)}=4.51$ MeV, $\bar{\eta}_n^{(2)}=2.46$, and a reduced proton SLD, $\bar{\eta}_p^{(2)}=0.64$, indicating a deformed proton shell contribution.
The third group, near $Q_{20}\approx20~\mathrm{b}$, has a much larger average deformation energy, $\bar{E}_{\mathrm{def}}^{(3)}=8.25$ MeV, but also small average SLD, $\bar{\eta}_n^{(3)}\simeq 1.84$ and $\bar{\eta}_p^{(3)}\simeq 0.83$, indicating deformed proton shell stabilization with an additional weaker neutron shell effect.
The coexistence of these three groups indicates that the light fragment is mainly driven by prolate proton shell stabilization around $Z=42$, with additional neutron shell effects emerging at larger quadrupole deformation.
Together with the near-spherical shell-stabilized heavy fragment, this light fragment structure completes the microscopic selectivity of the S1 channel.

Figure~\ref{fig:Pu240_S2_shell} shows the corresponding analysis for the S2 channel, represented by $^{144}$Ba [panels (a)--(c)] and $^{96}$Sr [panels (d)--(f)].
For the heavy fragment $^{144}$Ba ($Z=56$, $N=88$), the fragment ground state is located at finite octupole deformation, consistent with a pear-shaped configuration.
The scission configurations are located at sizable octupole deformation, $Q_{30}\approx2\text{--}2.5~\mathrm{b}^{3/2}$, and have a relatively large average deformation energy, $\bar{E}_{\mathrm{def}}=3.04$ MeV.
The averaged SLD, $\bar{\eta}_n=1.73$ and $\bar{\eta}_p=1.17$, are therefore larger than those of the near-spherical $^{134}$Te heavy fragment of the S1 channel.
Instead, the shell selectivity is associated with the overlap of the scission configurations with low $\eta_p$ and low $\eta_n$ regions with significant octupole deformation.
This pattern identifies an octupole deformed shell-stabilized configuration, with the main proton shell stabilization around $Z=56$ and an additional deformed neutron shell contribution around $N=88$.
For the complementary light fragment $^{96}$Sr ($Z=38$, $N=58$), the fragment ground state is located at finite $Q_{20}$ and small $Q_{30}$, corresponding to a mildly deformed shape.
The scission configurations have a relatively large average deformation energy, $\bar{E}_{\mathrm{def}}=4.54$ MeV, and overlap mainly with a low $\eta_p$ region, yielding $\bar{\eta}_p=1.12$ and indicating a deformed proton shell contribution.
By contrast, the averaged neutron SLD remains larger, $\bar{\eta}_n=1.97$, pointing to a weaker neutron shell contribution.
The S2 channel can therefore be interpreted as a deformed shell-stabilized configuration in which the octupole deformed shell structure of the heavy fragment provides the dominant microscopic selectivity, while the light fragment contributes mainly through a deformed proton shell effect around $Z=38$.

The comparison between Figs.~\ref{fig:Pu240_S1_shell} and~\ref{fig:Pu240_S2_shell} shows that the S1 channel exhibits more coherent microscopic selectivity, with the near-spherical $^{134}$Te fragment providing strong shell stabilization and $^{106}$Mo adding mainly prolate proton shell stabilization. By contrast, the S2 channel is driven by a more distributed deformed shell mechanism, dominated by the octupole deformed $^{144}$Ba fragment and supplemented by a weaker proton shell contribution from $^{96}$Sr. This difference provides a microscopic interpretation of the larger S1 yield and the smaller S2 yield.

\subsubsection{$^{232}$Th}

Figure~\ref{fig:Th232_S1_shell} shows the fragment PES,  $\eta_n$, and $\eta_p$ maps for the S1 channel, represented by $^{136}$Te [panels (a)--(c)] and $^{96}$Sr [panels (d)--(f)].
For the heavy fragment $^{136}$Te ($Z=52$, $N=84$), the fragment ground state is located at a spherical configuration.
The scission configurations are concentrated in a narrow band at finite octupole deformation, around $Q_{30}\approx0.7\text{--}1.0~\mathrm{b}^{3/2}$, rather than near the spherical ground state.
They occupy a relatively favorable region of the fragment PES at finite deformation energy, with $\bar{E}_{\mathrm{def}}=2.83$ MeV, and overlap with regions of low averaged SLD, $\bar{\eta}_n=0.67$ and $\bar{\eta}_p=0.02$.
This indicates that the heavy fragment $^{136}$Te retains neutron and proton shell stabilization associated with $Z=52$ and the nearby $N=82$ closure.
Compared with $^{134}$Te in the $^{240}$Pu S1 channel, the shift from $N=82$ to $N=84$ weakens the near spherical neutron shell stabilization, while the proton shell structure remains effective over a finite deformation region and supports the scission configurations at finite octupole deformation. 
For the complementary light fragment $^{96}$Sr, the scission configurations are concentrated in a compact region at finite deformation energy, with $\bar{E}_{\mathrm{def}}=3.61$ MeV, around $Q_{20}\approx10~\mathrm{b}$ and small $Q_{30}\approx1~\mathrm{b}^{3/2}$.
They correlate with a low $\eta_p$ region, with $\bar{\eta}_p=1.12$, whereas the averaged neutron SLD remains relatively large, $\bar{\eta}_n=2.05$, indicating a weaker neutron shell contribution.
This behavior, also found for the $^{96}$Sr light fragment in the S2 channel of $^{240}$Pu, points to a deformed proton shell contribution around $Z=38$.
Together, the weakened but still evident near-spherical shell stabilization of the $^{136}$Te and the deformed proton shell contribution of the $^{96}$Sr produce a relatively coherent microscopic selectivity in the S1 channel.

Figure~\ref{fig:Th232_S2_shell} shows the corresponding analysis for the S2 channel, represented by $^{140}$Xe [panels (a)--(c)] and $^{92}$Kr [panels (d)--(f)].
For the heavy fragment $^{140}$Xe ($Z=54$, $N=86$), the fragment ground state is located at a spherical configuration.
The scission configurations are located at finite octupole deformation, $Q_{30}\approx0.9~\mathrm{b}^{3/2}$, with an average deformation energy of $\bar{E}_{\mathrm{def}}=2.20$ MeV.
The averaged SLD, $\bar{\eta}_n=2.00$ and $\bar{\eta}_p=1.58$, are larger than those obtained for the $^{136}$Te heavy fragment in the S1 channel, indicating weaker shell selectivity.
The heavy fragment is therefore associated mainly with deformed proton shell stabilization around $Z=54$, but this effect is less pronounced than the cooperative stabilization in the S1 channel.
For the complementary light fragment $^{92}$Kr ($Z=36$, $N=56$), the fragment ground state is located at a spherical configuration.
The scission configurations are located at finite octupole deformation and have a large average deformation energy, $\bar{E}_{\mathrm{def}}=6.21$ MeV.
The average SLD remains relatively large, with $\bar{\eta}_n=1.68$ and $\bar{\eta}_p=2.20$, indicating a weak neutron shell contribution.
Thus, the light fragment exhibits weak microscopic selectivity and primarily serves as the complementary partner to the shell-stabilized heavy fragment in the S2 channel.

The comparison between Figs.~\ref{fig:Th232_S1_shell} and~\ref{fig:Th232_S2_shell} shows that S1 has more coherent microscopic selectivity, with weakened near-spherical stabilization in $^{136}$Te and a deformed proton shell contribution in $^{96}$Sr. By contrast, S2 shows weaker and more diffuse selectivity, mainly from a less pronounced deformed proton shell contribution in $^{140}$Xe, while $^{92}$Kr provides only weak additional stabilization.
This difference provides a microscopic interpretation of the larger S1 yield and the weaker, less distinctly separated S2 component in $^{232}$Th.

\subsubsection{$^{264}$Fm}

Figure~\ref{fig:Fm264_sym_shell} shows the fragment PES, $\eta_n$, and $\eta_p$ maps for the dominant symmetric channel represented by the $^{132}$Sn/$^{132}$Sn split.
The scission configurations are concentrated near the spherical ground state, where the low $\eta_n$ and low $\eta_p$ regions coincide, leading to small average deformation energy, and small average SLD for both neutron and proton.
This coherent overlap of the populated configurations with the fragment PES basin and low SLD regions reflects the spherical proton and neutron shell closures at $Z=50$ and $N=82$ in $^{132}$Sn.
Consequently, the dominant symmetric channel exhibits the clearest shell effect selectivity and is directly reflected in the yield maximum.

\begin{table}[b]
\caption{
Summary of the average deformation energy and average SLD for the representative fragments.
}
\label{tab:shell_summary}
\centering
\small
\setlength{\tabcolsep}{3.0pt}
\renewcommand{\arraystretch}{1.45}
\begin{ruledtabular}
\begin{tabular}{l c l c c c c c}
Nucleus & Ch. & Frag. & $\bar E_{\rm def}$ & $\bar{\eta}_n$ & $\bar{\eta}_p$ & \multicolumn{2}{c}{Shell} \\
        &     &       & (MeV)              &                  &                  & $N$ & $Z$ \\
\hline
$^{240}$Pu & S1 & $^{134}$Te      & 0.90 & $0.04$ & $0.00$ & 82 & 52 \\
           &    & $^{106}$Mo$^{(1)}$ & 0.94 & 2.18 & $1.33$ & 64 & 42 \\
           &    & $^{106}$Mo$^{(2)}$ & 4.51 & 2.46 & $0.64$ & 64 & 42 \\
           &    & $^{106}$Mo$^{(3)}$ & 8.25 & $1.84$ & $0.83$ & 64 & 42 \\
\cline{2-8}
           & S2 & $^{144}$Ba      & 3.04 & $1.73$ & $1.17$ & 88 & 56 \\
           &    & $^{96}$Sr       & 4.54 & 1.97 & $1.12$ & 58 & 38 \\
\hline
$^{232}$Th & S1 & $^{136}$Te      & 2.83 & $0.67$ & $0.02$ & 84 & 52 \\
           &    & $^{96}$Sr       & 3.61 & 2.05 & $1.12$ & 58 & 38 \\
\cline{2-8}
           & S2 & $^{140}$Xe      & 2.20 & 2.00 & $1.58$ & 86 & 54 \\
           &    & $^{92}$Kr       & 6.21 & $1.68$ & 2.20 & 56 & 36 \\
\hline
$^{264}$Fm & Sym. & $^{132}$Sn    & 0.34 & $0.02$ & $0.00$ & 82 & 50 \\
\end{tabular}
\end{ruledtabular}
\end{table}

\textbf{Summary of the fragment shell analysis.} 
Table~\ref{tab:shell_summary} summarizes the fragment shell analysis discussed above, listing the average deformation energy, $\bar{E}_{\mathrm{def}}$, and the average SLD, $\bar{\eta}_n$ and $\bar{\eta}_p$, for the representative fragments.

Overall, the SLD provides a classification of the shell stabilization.
The largest yield components are characterized by the most coherent correspondence among the populated Langevin scission configurations, favorable fragment PES regions, and low SLD regions.
In $^{240}$Pu and $^{232}$Th, the dominant S1 channels are driven by near-spherical $^{134,136}$Te heavy fragments, which are associated with low average SLD around $N=82$ and $Z=52$, together with deformed proton shell contributions from the light fragments around $Z=42$ in $^{106}$Mo for $^{240}$Pu and $Z=38$ in $^{96}$Sr for $^{232}$Th.
The S2 channels show more distributed shell selectivity: in $^{240}$Pu, $^{144}$Ba retains deformed neutron and proton shell stabilization around $N=88$ and $Z=56$ and is supplemented by a $Z=38$ proton shell contribution from $^{96}$Sr, whereas in $^{232}$Th the S2 channel shows only a modest proton contribution in $^{140}$Xe and weak selectivity in $^{92}$Kr.
For $^{264}$Fm, the symmetric channel is driven by spherical proton and neutron shell closures at $Z=50$ and $N=82$ in $^{132}$Sn.
Thus, proton shell effects provide the persistent microscopic selectivity in heavy and light fragments, whereas neutron shell effects reinforce this selectivity around $N=82$ and deformed configurations near $N=88$.

\section{Conclusion}
\label{Conclusion}

In this work, we investigated the SF of $^{240}$Pu, $^{232}$Th, and $^{264}$Fm to clarify the dynamical selection of fragment shell effects in the final yields.
% to clarify how collective dynamics select fragment shell effects in the final yields.
The calculation is based on a two-step microscopic framework: constrained HFB calculations provide the PES and collective inertia, and WKB tunneling connects the inner and outer turning points.
The subsequent Langevin dynamics describes the evolution through the classically allowed region to scission.
The resulting Langevin scission configurations are used both to construct the fragment yields and to provide the fragment deformations for the shell analysis.
The calculated yields display a two-component asymmetric pattern in $^{240}$Pu, a less distinctly separated asymmetric pattern in $^{232}$Th, and a symmetric peak in $^{264}$Fm. 

Fragment shell effects are reflected in the final SF yields only when they are dynamically accessible, with the PES topology providing the available valleys and the Langevin descent selecting the populated shell-favored scission configurations.
The yield peaks therefore reflect the combined selectivity of the Langevin dynamics and fragment shell stabilization.

In summary, the present framework offers a microscopic path for linking collective dynamics, fragment shell effects, and yield formation in SF.
Future work will extend this analysis to actinide and pre-actinide systems that exhibit rapid changes in fission modes.

\section*{DATA AVAILABILITY}

The data that support the findings of this article are openly available~\cite{data}.

\section*{Acknowledgments}

This work was supported by the Guangdong Basic and Applied Basic Research Foundation under Grant No.~2024A1515012310 and the National Natural Science Foundation of China under Grant No.~12305129 and No.~12475136.

\bibliography{aps-aichecked}% Produces the bibliography via BibTeX.

\end{document}